\documentclass[prc,twocolumn,10pt]{revtex4}

\usepackage{draftcopy}
\usepackage{epsfig}
\newcommand{\Version}       {Version: RatioDraft4.0.tex, July 12, 2002}
\newcommand{\bnl}           {$\rm^{1}$}
\newcommand{\ires}          {$\rm^{2}$}
\newcommand{\kraknuc}       {$\rm^{3}$}
\newcommand{\krakow}        {$\rm^{4}$}
\newcommand{\baltimore}     {$\rm^{5}$}
\newcommand{\newyork}       {$\rm^{6}$}
\newcommand{\nbi}           {$\rm^{7}$}
\newcommand{\texas}         {$\rm^{8}$}
\newcommand{\bergen}        {$\rm^{9}$}
\newcommand{\bucharest}     {$\rm^{10}$}
\newcommand{\kansas}        {$\rm^{11}$}
\newcommand{\oslo}          {$\rm^{12}$}

\begin{document}

\title{Rapidity Dependence of Charged Antiparticle-to-Particle Ratios in
Au+Au Collisions at $\sqrt{s_{NN}}=200$ GeV}

\date{\Version}

\author{
  I.~G.~Bearden\nbi, 
  D.~Beavis\bnl, 
  C.~Besliu\bucharest, 
  Y.~Blyakhman\newyork, 
  B.~Budick\newyork, 
  H.~B{\o}ggild\nbi, 
  C.~Chasman\bnl, 
  C.~H.~Christensen\nbi, 
  P.~Christiansen\nbi, 
  J.~Cibor\kraknuc, 
  R.~Debbe\bnl, 
  E. Enger\oslo,  
  J.~J.~Gaardh{\o}je\nbi, 
  M.~Germinario\nbi, 
  K.~Hagel\texas, 
  O.~Hansen\nbi, 
  A.~Holm\nbi, 
  A.~K.~Holme\oslo, 
  H.~Ito\kansas, 
  E.~Jakobsen\nbi, 
  A.~Jipa\bucharest, 
  F.~Jundt\ires, 
  J.~I.~J{\o}rdre\bergen, 
  C.~E.~J{\o}rgensen\nbi, 
  R.~Karabowicz\krakow, 
  T.~Keutgen\texas, 
  E.~J.~Kim\bnl, 
  T.~Kozik\krakow, 
  T.~M.~Larsen\oslo, 
  J.~H.~Lee\bnl, 
  Y.~K.~Lee\baltimore, 
  G.~L{\o}vh{\o}iden\oslo, 
  Z.~Majka\krakow, 
  A.~Makeev\texas, 
  B.~McBreen\bnl, 
  M.~Mikelsen\oslo, 
  M.~Murray\texas, 
  J.~Natowitz\texas, 
  B.~S.~Nielsen\nbi, 
  J.~Norris\kansas, 
  K.~Olchanski\bnl, 
  J.~Olness\bnl, 
  D.~Ouerdane\nbi, 
  R.~P\l aneta\krakow, 
  F.~Rami\ires, 
  C.~Ristea\bucharest, 
  D.~R{\"o}hrich\bergen, 
  B.~H.~Samset\oslo, 
  D.~Sandberg\nbi, 
  S.~J.~Sanders\kansas, 
  R.~A.~Scheetz\bnl, 
  P.~Staszel\nbi, 
  T.~S.~Tveter\oslo, 
  F.~Videb{\ae}k\bnl, 
  R.~Wada\texas, 
  A.~Wieloch\krakow, 
  Z.~Yin\bergen, 
  I.~S.~Zgura\bucharest\\ 
  The BRAHMS Collaboration \\ [1ex]
  \bnl~Brookhaven National Laboratory, Upton, New York 11973, USA\\
  \ires~Institut de Recherches Subatomiques and Universit{\'e} Louis
  Pasteur, Strasbourg, France\\
  \kraknuc~Institute of Nuclear Physics, Krakow, Poland\\
  \krakow~Smoluchkowski Inst. of Physics, Jagiellonian University, Krakow, Poland\\
  \baltimore~Johns Hopkins University, Baltimore 21218, USA\\
  \newyork~New York University, New York 10003, USA\\
  \nbi~Niels Bohr Institute, Blegdamsvej 17, University of Copenhagen, Copenhagen 2100, Denmark\\
  \texas~Texas A$\&$M University, College Station, Texas, 17843, USA\\
  \bergen~University of Bergen, Department of Physics, Bergen, Norway\\
  \bucharest~University of Bucharest, Romania\\
  \kansas~University of Kansas, Lawerence, Kansas 66049, USA \\
  \oslo~University of Oslo, Department of Physics, Oslo, Norway\\
 }

\begin{abstract}
  We present ratios of the numbers of charged antiparticles to
  particles (pions, kaons and protons) in Au + Au collisions at
  $\sqrt{s_{NN}}=200$ GeV as a function of rapidity in the range
  $y$=0-3. While the particle ratios at midrapidity are approaching unity, the
  $K^-/K^+$ and $\bar{p}/p$ ratios decrease significantly at forward
  rapidities. An interpretation of the results within the statistical
  model indicates a reduction of the baryon chemical
  potential from $\mu_B \approx 130$MeV at $y$=3 to $\mu_B
  \approx 25$MeV at $y$=0.\\

\end{abstract}

\maketitle



Ratios of yields of particles to antiparticles and, in particular
the rapidity dependence of such ratios, are significant indicators
of the dynamics of high energy nucleus-nucleus
collisions~\cite{Hermann99,Satz2000}. At the energy of
$\sqrt{s_{NN}}$=200 GeV
considerable transparency is expected for Au+Au collisions, even
for central events. This should lead to a flat multiplicity
density as function of rapidity ($y$) near midrapidity, and
$\bar{p}/p$ and $K^{-}/K^{+}$ particle yield ratios with values
near unity. Away from midrapidity the net baryon content
originating from the initial nuclei is significant and production
mechanisms other than particle-antiparticle pair production play a
substantial role. Therefore $\bar{p}/p$ and $K^{-}/K^{+}$ ratios
will decrease with increasing rapidity $|y|$. Measurements of
$\bar{p}/p$ ratios at 130
GeV~\cite{BRAHMSratio130,STARratio130,PHOBOSratio130,PHENIXratio130}
and of pseudo-rapidity distributions of charged particles at 130
GeV~\cite{BRAHMSmult130} and 200
GeV~\cite{BRAHMSmult200,PHOBOSmult200} point to the development
described above, reminiscent of the Bjorken
picture~\cite{Bjorken83}.

In the present Letter measurements of ratios $\pi^{-}/\pi^{+},
K^{-}/K^{+}$ and $\bar{p}/p$ as a function of rapidity, transverse
momentum ($p_{T}$) and collision centrality (top 20\%) are
presented for Au+Au collisions at $\sqrt{s_{NN}}=200$ GeV. We
measure $\langle d^{2}n/dp_{T}dy\rangle \Delta p_{T} \Delta y$,
where the averaging is over phase space, $\Delta p_{T} \Delta y$.
The ratios are for the normalized numbers of detected particles,
including corrections as described below.

At midrapidity, the measured antiparticle to particle ratios are
approaching unity, i.e.\ they vary between $0.75\pm 0.04$
($\bar{p}/p$), $0.95\pm 0.05$ ($K^-/K^+$) to $1.01 \pm 0.04$
($\pi^-/\pi^+$). These values exceed the corresponding ratios
measured for Pb+Pb reactions at ($\sqrt{s_{NN}}=17$ GeV) by
factors of about 1.8 for the kaons and by about 10 for the
protons. Recent PHOBOS measurements~\cite{PHOBOS200ratio} are
consistent with the $y$=0 measurements presented here.

While the pion ratios are consistent with unity over the entire
rapidity range covered, the ratios of the strange mesons and of
the baryons drop to $K^-/K^+= 0.67\pm 0.06 $ and $\bar{p}/p =
0.23\pm0.03$ at $y \approx 3$.  This behavior supports the idea of
the formation of a nearly net-baryon free zone at midrapidity in
the collisions where particle production is dominated by pair
creation, either at the quark level or at the hadron level.
However, towards larger rapidities, the ratios become increasingly
influenced by the baryon content of the original nuclei.  The
present work indicates an increase, at $y=0$, of the ratios
relative to $\sqrt{s_{NN}}=130$ GeV Au+Au collisions of about 5\%
for kaons and of about 17\% for protons
~\cite{BRAHMSratio130,STARratio130,PHOBOSratio130,PHENIXratio130}.

The data were obtained with the BRAHMS detector, which consists of
two independent small-aperture magnetic spectrometers that can
rotate in the horizontal plane about the nominal interaction point
(IP), covering the rapidity range $-0.1< y < 4$ for pions and
$-0.1< y < 3.4$ for protons. Details may be found in
Ref.~\cite{BRAHMSNIM}. The MidRapidity Spectrometer (MRS) consists
of two time projection chambers (TPC) and a magnet, for
determining particle momenta. This assembly is followed by a
segmented scintillator time--of--flight wall (TOFW), with time
resolution $\sigma_t \approx 75$ps, for particle velocity
measurements. Requiring a $\pm 2\sigma_t$ cut around the expected
flighttime, $\pi$-$K$ separation is achieved up to a momentum of
2.3 GeV/c and $K$-$p$ separation up to 3.9 GeV/c. The Front
Forward Spectrometer (FFS) consists, in order, of a dipole magnet,
a TPC, a second dipole magnet, a second TPC, a time--of--flight
wall (TOF1) and a threshold gas-Cherenkov detector (C1). At small
polar angles (from $2.3^\circ$ to $15^\circ$), where the mean
momentum of particles is large, a back section (BFS) with two
dipole magnets, three drift chambers, a time of flight wall (TOF2)
and a ring imaging Cherenkov detector (RICH) is also used. TOF1
(at 8.6 m) and TOF2 (at 18 m) allow for $K$-$p$ separation up to
$p$=5.5 and 8 GeV/c, respectively. C1 identifies pions in the
range from $p=3$ to 9 GeV/c and the RICH allows $\pi$-$K$
separation up to $p=25$ GeV/c and $K$-$p$ separation from $p$=10
to $p$=35 GeV/c.

\begin{figure}[htp]
  \epsfig{file=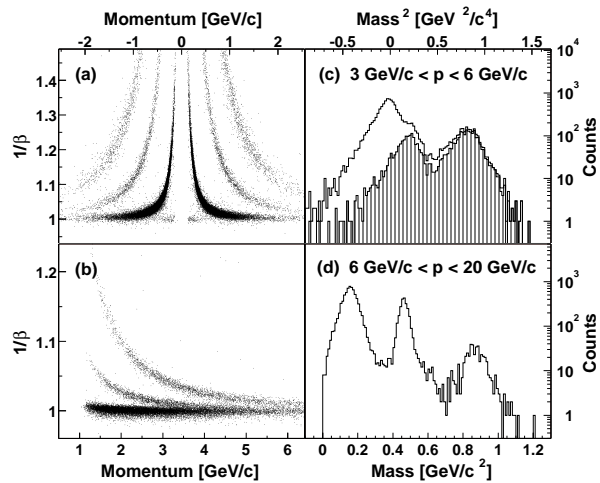,width=\linewidth}
  \caption{ Particle identification capability in BRAHMS. (a) and (b)
    Separation between pions, kaons, and protons. The MRS (a) is at 90
    deg.\ and both charges are accepted. The FS (b) is here positioned
    at 12 degrees. (c) Mass-squared spectrum in the FS from
    time-of-flight measurements. The shaded histogram includes the
    vetoing of pions in C1. (d) Mass spectrum in FS using the RICH.}
 \label{fig1}
 \vspace{-3mm}
\end{figure}

The reaction centrality was determined using a plastic scintillator
tile array surrounding the intersection
region~\cite{BRAHMSNIM,BRAHMSmult130,BRAHMSmult200}.
Beam-Beam Counters (BBCs) consisting of two arrays of Cherenkov
radiators positioned $\pm$~2.15 m from the IP were used to measure
charged hadrons in the pseudo-rapidity range $3.0<|\eta|<3.8$. For
the most $25\%$ central collisions, the BBCs allow collision
vertex determination with resolution $\sigma_z \approx 0.65 $ cm
and supply the start time for the time--of--flight measurements
with $\sigma_t \lesssim 30$ ps.

The data presented here were collected with the MRS at $40^\circ$,
$60^\circ$, and $90^\circ$ and the FS at
$4^\circ$,$8^\circ$,$12^\circ$ and $20^\circ$.  For magnetic
fields of the same magnitude but opposite polarities, the
spectrometer acceptance is identical for positively and negatively
charged particles. Therefore, most systematic errors associated
with acceptance and detector efficiency cancel in the particle
ratios. In general, the experimental methods employed here are the
same as those described in~\cite{BRAHMSratio130}.

\begin{figure*}[htp]
  \epsfig{file=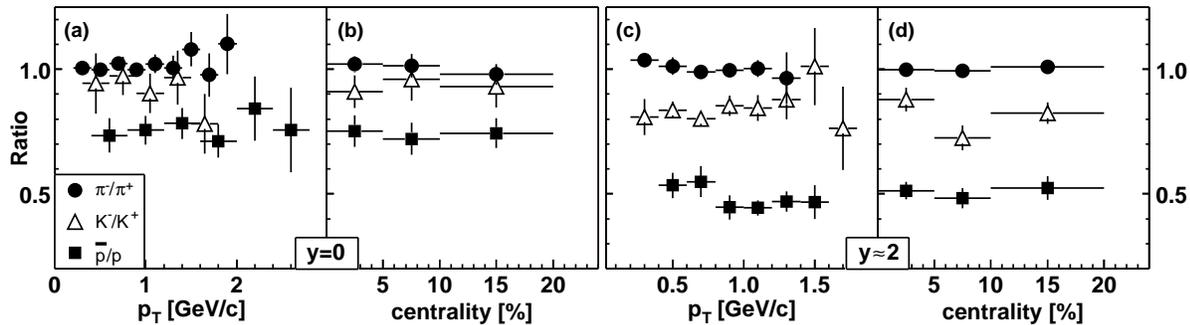,width=\linewidth}
  \caption{ Dependence of the measured antihadron/hadron ratios
    on transverse momentum and reaction centrality. The panels to the
    left are for $y=0$ and panels to the right for
    rapidities in the range $y=2.05$ for protons to $y=2.55$ for
    pions.The errors are statistical only. }
  \label{fig2}
  \vspace{-4mm}
\end{figure*}

Figure~\ref{fig1} shows the particle identification capability.
Particle yields in the MRS and FS are determined by selecting
tracks within a $\pm 2\sigma$ band of the expected $\beta^{-1}$
vs. momentum behavior for a given particle type. In the FS, additional
momentum dependent cuts on the Cherenkov detector signals have
been applied. The number of particles measured in various runs is
normalized to the number of collision events fulfilling a given
centrality cut. The geometrical acceptances of the spectrometers
change as a function of the collision vertex and the
normalization is done depending on
the collision vertex, thus accounting for possible differences in
the vertex distributions from run to run. From the normalized
number of antiparticles and particles the ratios are calculated.

The ratios have been corrected for absorption and production of
secondary particles in the material traversed (Be beam tube,
various detector elements and air). Losses of antiprotons due to
annihilation have been evaluated in GEANT simulations to be about
3\% in the MRS and 0.9\% in the FS. In the MRS the background
contribution to proton yields, arising mainly from the interaction
of pions with the beam pipe, is $\approx 10\%$ for the lowest
$p_{T}$ bin at midrapidity falling off to 2 \% at $p_T > 0.5$GeV.
In the FS this contribution is found to be negligible at the
smallest angles and amounts to less than 1\% at 30 degrees. For
kaons decay losses cancel out, and differences in reaction cross
sections for $K^-$ and $K^+$ amount to less than 1\%.

Figure~\ref{fig2} shows the dependence of the ratios for pions,
kaons and protons (corrected as described above) on transverse
momentum $p_T$ (a and c) and collision centrality (b and d) in two
selected rapidity intervals, around $y$=0 and $y\approx$2
respectively.  The measured ratios show no significant dependence
on $p_T$ or centrality in the covered ranges. Therefore we
integrate our yields over centrality in the 0-20\% range and over
transverse momentum before calculating the ratios.

The ratios shown in Fig.~\ref{fig2} have not been corrected for
protons and antiprotons that originate from weak decays of
hyperons ($\Lambda, \Sigma$, $etc.$). Corrections for feed down
depend on the relative production of hyperons and primary baryons
and their antiparticles, and on the respective spectrum slopes.
The STAR experiment has recently measured $\Lambda / p \approx
0.5$ and $\bar{\Lambda} / \Lambda$ = $0.74 \pm 0.04$ at $y$=0 for
$\sqrt{s_{NN}}=130$ GeV collisions~\cite{STARlambda130}. We have
studied the magnitude of the corrections using various model
assumptions as input to realistic GEANT simulations of the BRAHMS
setup.  Assuming primary Hyperon/Baryon ratios similar to those
measured by STAR we find that the corrections to the quoted ratios
in our acceptance are less than 5\%.

\begin{figure}[htp]
  \vspace{-4mm}
  \epsfig{file=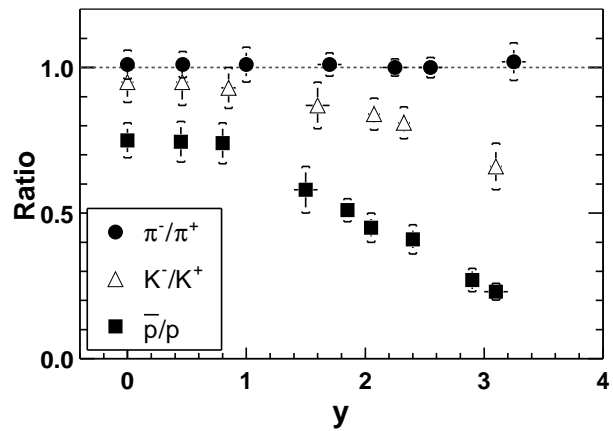,width=\linewidth}
  \vspace{-8mm}
  \caption{Antiparticle-to-particle ratios as a function of
    rapidity. The vertical lines show the statistical errors while the
    caps indicate the combined statistical and systematic errors.}
  \label{fig3}
\end{figure}

Figure~\ref{fig3} shows the $\pi^{-}/\pi^{+}$, $K^{-}/K^{+}$ and
$\bar{p}/p$ ratios as a function of rapidity. Statistical errors
are shown as vertical error bars, while systematic plus
statistical errors are indicated by the caps.
Systematic uncertainties are estimated as 4\% primarily from the
normalizations between opposite polarity settings. While the
$\pi^{-}/\pi^{+}$ ratio is consistent with unity over the
considered rapidity range, the $K^{-}/K^{+}$ ratio shows a
decrease from $0.95\pm0.05$, at $y=0$, to $0.67\pm0.06$, at
$y=3.05$.  Likewise the $\bar{p}/p$ ratio decreases from
$0.75\pm0.04$, at $y=0$, to $0.23 \pm 0.03$, at $y=3.1$. The
$\bar{p}/p$ ratio at $y$=0 exceeds the ratio measured in Au+Au
collisions at 130
GeV~\cite{BRAHMSratio130,PHOBOSratio130,STARratio130} by about
17\%. We also note that the $\bar{p}/p$ and $K^{-}/K^{+}$ ratios
are essentially constant in the rapidity interval $y$=0-1. This is
consistent with the onset of the boost invariant plateau around
midrapidity.

The measured set of particle ratios at midrapidity also lends itself to an
analysis in terms of a model based on the assumption of a system
in chemical and thermal equilibrium such as has already been
advocated for SPS energies. Recently, particle ratios measured at
$\sqrt{s_{NN}}=130$ GeV in the midrapidity region have been
analyzed in a Grand Canonical Ensemble (GCE) with baryon number,
strangeness and charge conservation \cite{PBM01statmodel}. The
relevant parameters of the model are the temperature T and the
baryon (or light quark) chemical potential $\mu_B= 3\mu_q$. Values
of $T= 174 \pm 7$ MeV and $\mu_B= 46 \pm 5$ MeV were found. In
Ref.~\cite{PBM01statmodel} a parametrization as a function of
energy is proposed leading to a prediction for $\sqrt{s_{NN}} =
200$ GeV of $T=177 \pm 7$ MeV, $\mu_B= 29 \pm 8$ MeV and thus
$\bar{p}/p$ = 0.752, $K^-/K^+$= 0.932, and $\pi^-/\pi^+$= 1.004.
We note the excellent numerical agreement between these
calculations and the present measurements. Within this approach,
the near constancy of the temperature for chemical freeze-out
found at SPS, at lower RHIC energies, and at the present energy
can be associated with a fixed deconfinement transition
temperature and the establishment of chemical equilibrium during
hadronization. The small value of the chemical potential indicates
a small net baryon density at midrapidity.

\begin{figure}[htp]
  \epsfig{file=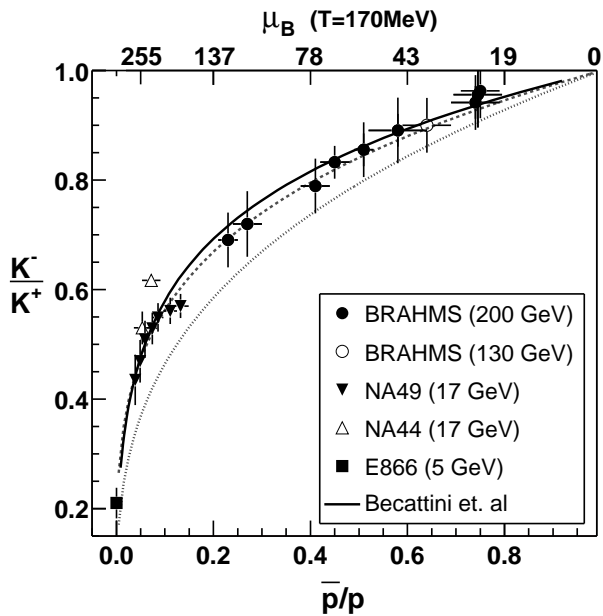,width=\linewidth}
  \vspace{-5mm}
  \caption{Correlation between strange meson and baryon antiparticle-to-particle ratios.
    The NA44 and NA49 data are from Refs.~\cite{NA44} and \cite{NA49},
    respectively. The E866 data are from Ref.~\cite{AGS}.  The dashed
    line shows the relation $K^-/K^+ = (\bar{p}/p)^{1/4}$ and the
    dotted line shows $K^-/K^+ = (\bar{p}/p)^{1/3}$. The full line shows the
    thermal prediction of Becattini et al.\cite{Becattini}.}
  \label{fig4}
\end{figure}

Surprisingly, a comparison of the $K^-/K^+$ and $\bar{p}/p$ ratios
over a large range in rapidity and collision energy shows a
remarkable correlation as shown in Fig.~\ref{fig4}. Ratios are
measured at slightly different rapidities. Therefore the $K^-/K^+$
ratios in Fig.~\ref{fig4} are interpolated to the $y$ value of the
$\bar{p}/p$ measurements. Also shown are similar ratios determined
at AGS~\cite{AGS} and SPS energies~\cite{NA49, NA44}. The figure
evidences a smooth relationship between the ratios from AGS and
SPS to RHIC, that can be expressed by a power law $K^-/K^+ =
(\bar{p}/p)^{1/4}$.
Since the compared ratios are not at the same rapidity a thermal
interpretation is not strictly justified. Nonetheless, we note that
for a
vanishing strange-quark chemical potential this exponent is
expected  to have a value of 1/3. The present result suggests
$\mu_s = 1/4 \times \mu_q$. Good agreement is also found by
comparing the present data with the thermal model of Becattini et
al.~\cite{Becattini}, which is based on SPS results integrated
over the full phase space,and assuming $T=170$MeV. Within the
framework of the statistical model, and assuming that the particle
sources corresponding to the different rapidity regions sampled in
our experiments are all in local chemical equilibrium subject to
strangeness conservation, Fig.~\ref{fig4} therefore suggests that
the baryon chemical potential decreases from $\mu_B \approx 130$
MeV, at $y\approx$3, to $\mu_B \approx 25$ MeV, at $y$=0.

In summary, the BRAHMS experiment has measured the ratios of
charged hadrons at the RHIC top energy and observed the highest
such ratios yet in nuclear collisions.  We find an increase in the
ratio of antiprotons to protons and charged singly-strange mesons,
consistent with a significant increase in reaction transparency
compared to lower energies.  The ratios are, within errors,
constant in the interval $y=0-1$ as expected for a boost invariant
midrapidity plateau dominated by particle production from the
color field. The ratios are well reproduced by statistical model
calculations in which the baryon chemical potential decreases
strongly from forward rapidities, where the baryon content of the
original colliding nuclei is still significant, towards
midrapidity. The systematics of kaon and proton ratios suggests a
empirical universal relationship between light and strange quark
chemical potentials.

This work was supported by the division of Nuclear Physics of the
Office of Science of the U.S. DOE, the Danish Natural Science
Research Council, the Research Council of Norway, the Polish State
Com. for Scientific Research and the Romanian Ministry of
Research. We are indebted to F. Becattini for supplying us with
thermal model calculations.

\newpage
\end{document}